\documentclass{PoS}

\title{Theory of rare charm decays into leptons}

\ShortTitle{Theory of rare charm decays into leptons}

\author{\speaker{Alexey A Petrov}%
         \thanks{This work has been supported in part by the U.S. Department of Energy under contract DE-SC0007983.}\\
         Department of Physics and Astronomy\\
	Wayne State University\\
	Detroit, MI 48201, USA \\
        E-mail: \email{apetrov@wayne.edu}}
        
\abstract{Studies of rare decays of charmed mesons into the final states containing leptons 
is an important vehicle in low-energy searches for new physics. I review theoretical implications of those
studies involving transitions with and without lepton flavor conservation. I argue that lepton-flavor violation (LFV) could be 
successfully studied in heavy quarkonium transitions yielding more sensitive results than similar searches  involving
$D$ mesons.}

\FullConference{9th International Workshop on the CKM Unitarity Triangle\\
		28  November - 3 December 2016\\
		Tata Institute for Fundamental Research (TIFR), Mumbai, India}

\begin{document}

%%%%%%%%%%%%%%%%%%%%%
\section{Introduction}\label{Section1}

Studies of flavor-changing neutral current (FCNC) transitions with charm quarks is an important vehicle in 
low-energy searches for new physics (NP) \cite{Intro}. The primary reason for it is the fact that the standard 
model (SM) Lagrangian does not contain terms that allow for a change of quark or lepton flavor while 
conserving their electric charge. Physical processes that represent FCNC transitions are, however, possible 
in the SM due to quantum fluctuations, i.e. by considering electroweak interactions at one loop. If NP interactions 
are such that FCNC transitions are possible, either due to elementary interactions in NP Lagrangian or via loop effects with new 
particles \cite{Petrov:2016azi}, studies of FCNC decays can prove useful in constraining properties of new physics states.  

FCNC transition rates are not necessarily small in the SM. Due to the left-handed nature of weak interactions, such 
currents would be induced with the coefficients that are proportional to the masses (squared) of quarks running in 
the electroweak loop diagrams generating FCNC transitions. They {\it are} expected be small in $D$-decays both 
due to relatively small mass of the intermediate bottom quark and tiny values of corresponding Cabbibo-Kobayashi-Maskawa (CKM) 
matrix elements. While some enhancements are possible due to long-distance QCD effects, transitions rates for rare decays 
such as $D^0 \to \ell^+ \ell^-$ are still small and have never been observed. This fact makes them a prime target for new 
physics searches in low energy experiments. Decays of charmed states can probe a variety of NP scenarios due to availability 
of large datasets of charmed particle decays. 

In what follows I shall review the theoretical status of leptonic rare decays of charmed meson states. I shall
argue that such decays can probe both quark-flavor violating (QFV) and lepton-flavor violating (LFV) transitions.
Since both up-quark and lepton FCNC transitions are generally expected to be small, it would be reasonable to consider 
QFV and LFV transitions separately.

%%%%%%%%%%%%%%%%%%%%%
\section{Lepton flavor conserving rare decays}\label{Section2}

Lepton-flavor conserving rare decays of $D$ mesons are mediated by quark-level transitions
$c \to u \bar \ell \ell$ and $c \to u \gamma^*$ (followed by $\gamma^* \to \bar\ell\ell$). Due to the hierarchical 
structure of the CKM matrix, the Glashow-Iliopoulos-Maiani  (GIM) mechanism is very effective in these 
transitions, which makes corresponding SM branching ratios small  \cite{Intro}.

In the SM, the short distance contribution to $c \to u \bar \ell \ell$ and $c \to u \gamma^*$ follows from the 
Lagrangian \cite{Intro}
\begin{equation}\label{SMLagrangian}
{\cal L}_{\rm eff}  = 
\frac{4 G_F}{\sqrt{2}} \sum_{i=7,9,10} C_i O_i,
\end{equation}
where $G_F$ is the Fermi constant and the effective operators are defined as 
\begin{eqnarray}
O_7 &=& -\frac{g_{em}}{16\pi^2} m_c (\overline{\ell}_L \gamma_\mu \ell_L) 
(\overline{u}_L \gamma^\mu
c_L)\ , 
\\
O_9 &=& \frac{\alpha}{4\pi}(\overline{\ell} \gamma_\mu \ell) 
(\overline{u}_L \gamma^\mu
c_L)\ , \ 
O_{10} = \frac{\alpha}{4\pi}(\overline{\ell} \gamma_\mu \gamma_5 \ell) 
(\overline{u}_L \gamma^\mu
c_L)\ .
\nonumber
\end{eqnarray}
where, numerically, the Wilson coefficients are $C_9(\mu = m_c) = 0.198 |V_{ub}V_{cb}|$,
$C_{10}(\mu = m_c) \simeq 0$ and $C_7^{\rm eff}(\mu =m_c)=-0.0025$ for $m_c = 1.3$ GeV
(see \cite{deBoer:2016dcg} for details).

All possible heavy NP contributions that generate $c \to u \ell^+ \ell^-$  transitions
can be summarized in terms of an effective Lagrangian ${\cal L}_{\rm NP}^{rare}$. 
\begin{equation}\label{NPLagrangian1}
{\cal L}_{\rm NP}^{rare}  = 
-\sum_{i=1}^{10}  \frac{{\rm \widetilde C}_i (\mu)}{\Lambda^2} ~ \widetilde Q_i,
\end{equation}
where ${\rm \widetilde C}_i$ are the Wilson coefficients, $ \widetilde Q_i$ are the effective operators, and
$\Lambda$ represents a scale of possible new physics interactions at which $\widetilde Q_i$ are generated.
There are only ten of those operators of dimension six, 
\begin{eqnarray}
\begin{array}{l}
\widetilde Q_1 = (\overline{\ell}_L \gamma_\mu \ell_L) 
(\overline{u}_L \gamma^\mu
c_L)\ , \\
\widetilde Q_2 = (\overline{\ell}_L \gamma_\mu \ell_L)  
(\overline{u}_R \gamma^\mu
c_R)\ , \\ 
\widetilde Q_3 = (\overline{\ell}_L \ell_R) \ (\overline{u}_R c_L) \ , 
\end{array}
\qquad 
\begin{array}{l}
\widetilde Q_4 = (\overline{\ell}_R \ell_L) 
(\overline{u}_R c_L) \ , \\
\widetilde Q_5 = (\overline{\ell}_R \sigma_{\mu\nu} \ell_L) 
( \overline{u}_R \sigma^{\mu\nu} c_L)\ ,\\
\phantom{xxxxx} 
\end{array}
\label{SetOfOperatorsLL}
\end{eqnarray}
where five additional operators $\widetilde Q_6, \dots, \widetilde Q_{10}$ 
that can be obtained from operators in Eq.~(\ref{SetOfOperatorsLL}) by 
interchanging left- and right-handed fields (i.e. switching $L \leftrightarrow R$), e.g. 
$\widetilde Q_6 =  (\overline{\ell}_R \gamma_\mu \ell_R) (\overline{u}_R \gamma^\mu c_R)$,
$\widetilde Q_7 =  (\alpha/4) (\overline{\ell}_R \gamma_\mu \ell_R) (\overline{u}_L \gamma^\mu c_L)$, etc.

The effective Lagrangian of Eq.~(\ref{SetOfOperatorsLL}) leads to lepton-flavor conserving QFV transitions. 
Of those, the simplest one is a rare decay $D^0 \to \ell^+ \ell^-$. Not all operators from Eq.~(\ref{SetOfOperatorsLL})  
contribute to the decay rate, as some matrix elements (or their linear combinations) vanish in 
the calculation of ${\cal B} (D^0\to \ell^+\ell^-)$. For instance, $\langle \ell^+ \ell^- | \widetilde Q_5 | D^0 \rangle =
\langle \ell^+ \ell^- | \widetilde Q_{10} | D^0 \rangle = 0$ (quantum numbers mismatch),
$\langle \ell^+ \ell^- | Q_9 | D^0 \rangle \equiv  (\alpha/4) \langle \ell^+ \ell^- | (\widetilde Q_1 + \widetilde Q_7) | D^0 \rangle = 0$ 
(vector current conservation), etc. This transition has a very small SM contribution \cite{Intro}, so it could serve as a clean 
probe of amplitudes induced by NP particles. The most general $D^0 \to \ell^+ \ell^-$ decay amplitude is given by
\begin{equation}\label{decayampl}
{\cal A} (D^0\to \ell^+\ell^-) = {\overline u}(p_-, s_-) 
\left[ \ A + \gamma_5 B \ \right] v(p_+, s_+), 
\end{equation}
where ${\overline u}(p_-, s_-)$ and $v(p_+, s_+)$ are leptons' spinors.
The constants $A$ and $B$ depend on the Wilson coefficients of the Lagrangian Eq.~(\ref{NPLagrangian1}) and
some hadronic parameters,
\begin{eqnarray}
\left| A\right|  &=& \frac{f_D M_D^2}{4 \Lambda^2 m_c} \left[\widetilde C_{3-8} + 
\widetilde C_{4-9}\right]\ , \nonumber 
 \\
\left| B\right|  &=& \frac{f_D}{4 \Lambda^2} \left[
2 m_\ell \left(\widetilde C_{1-2} + \widetilde C_{6-7}\right)
+  \frac{M_D^2}{m_c}
\left(\widetilde C_{4-3} + \widetilde C_{9-8}\right)
\right]\ ,  \label{DlCoeff}
\end{eqnarray}
with $\widetilde C_{i-k} \equiv \widetilde C_i-\widetilde C_k$. The amplitude of Eq.~(\ref{decayampl})
results in the branching fraction,
\begin{equation}\label{Dllgen}
{\cal B} (D^0 \to \ell^+\ell^-) = 
\frac{M_D}{8 \pi \Gamma_{\rm D}} \sqrt{1-\frac{4 m_\ell^2}{M_D^2}}
\left[ \left(1-\frac{4 m_\ell^2}{M_D^2}\right)\left|A\right|^2  +
\left|B\right|^2 \right] \ \ , 
\end{equation}
According to Eq.~(\ref{DlCoeff}), the standard model contribution vanishes in the $m_\ell \to 0$ limit. 
Any NP model that contribute to $D^0 \to \ell^+ \ell^-$ can be constrained
from the bounds on the Wilson coefficients in Eq.~(\ref{DlCoeff}). It is important to
point out that because of the helicity suppression, studies of ${\cal B} (D^0 \to e^+ e^-)$ vs. 
${\cal B} (D^0 \to \mu^+\mu^-)$ (and therefore analyses of lepton universality in those decays) are very 
complicated experimentally (see Sec. \ref{Section4} for more on this). 

Experimental studies of  $D^0 \to \ell^+ \ell^-$ transitions result in the 
upper bounds on the branching fractions \cite{Amhis:2016xyh,Aaij:2013cza},
\begin{equation}
{\cal B} (D^0 \to \mu^+\mu^-) <  7.6\times 10^{-9}, \ \
{\cal B} (D^0 \to e^+ e^-) < 7.9\times 10^{-8}, 
\label{brs}
\end{equation}
In studying NP contributions to rare decays in charm, it might be advantageous to 
study {\it correlations} of various processes, for instance
$D^0\overline{D}^0$ mixing and rare decays~\cite{Golowich:2009ii}.

The effective operators that do not contribute to $D^0 \to \ell^+ \ell^-$ can be studied in rare
semileptonic \cite{deBoer:2015boa,Fajfer:2007dy} and radiative leptonic $D$-decays. 
Finally, similar decays with neutrino final states can be used to constrain not only
possible NP contributions to $c \to u \nu \bar\nu$ \cite{Burdman:2001tf}, but also NP models with 
light dark matter particles \cite{Badin:2010uh}, as those transitions have the same experimental signature.

%%%%%%%%%%%%%%%%%%%%%
\section{Lepton flavor violating rare decays}\label{Section3}

Lepton flavor does not need to be conserved. In the standard model the FCNC interactions
in the lepton sector are small, as flavor-violating transition is proportional to square of neutrino masses.
This fact makes the rates of LFV FCNC interactions tiny in teh SM. There are, however, many new physics models where 
lepton flavor is also not conserved. This makes studies of LFV a background-free search for such models of NP.

The discussion of lepton-flavor conserving rare decays in Sect. \ref{Section2} could be easily extended to 
include LFV transitions. For example, the rate of LFV leptonic decays of $D$ would be,
\begin{equation}\label{Dllg2}
{\cal B} (D^0 \to \mu^+e^-) = 
\frac{M_D}{8 \pi \Gamma_{\rm D}} 
\left( 1-\frac{ m_\mu^2}{M_D^2} \right)^2 
\left[ \left|A\right|^2  + \left|B\right|^2 \right] \ \,
\end{equation}
where I neglected the electron mass. The constants $A$ and $B$ are
defined in Eq.~(\ref{decayampl}). Experimental bounds on LFV decays 
exist \cite{Amhis:2016xyh}, e.g. 
\begin{equation}
{\cal B} (D^0 \to \mu^\pm e^\mp) <  1.3\times 10^{-8}.
\label{brs1}
\end{equation}
It is important to notice that the LFV decay in Eq.~(\ref{Dllg2}) involves FCNC transitions {\it twice}, both on quark and 
lepton sides of the effective operator that generates such interaction. While this is indeed possible, such operators must be additionally 
suppressed in many NP models compared to the similar lepton-flavor conserving transitions, say of Eq.~(\ref{Dllgen}). 
Thus, if we insist on studying LFV transitions in meson or baryon decays, we should require flavor conservation on the quark side. 
This immediately implies that Eq.~(\ref{Dllg2}) and similar ones will only probe the operators of the type
$\left(\bar \mu e\right)\left(\bar u u\right)$, $\left(\bar \mu e\right)\left(\bar c c\right)$, or 
$\left(\bar \mu e\right)\left(\bar b b\right)$. All of those decays are suppressed by small combinations of 
the CKM factors $V_{cb}^* V_{ub}$ resulting from quark FCNCs in the SM in $c\to u \bar \mu e$. It is 
thus more beneficial to probe those operators in the decays that do not involve quark FCNC interactions, such
as two-body $b\bar b$, $c\bar c$, or $u\bar u$ quarkonium decays \cite{Hazard:2016fnc}.
This argument can be easily extended for other LFV transitions of the types $D \to \bar \ell_1 \ell_2$ or 
$D \to M \bar \ell_1 \ell_2$ where $M$ is a light-quark meson. 

The effective Lagrangian describing such transition, ${\cal L}_{\rm eff}$, can then be divided into the dipole part, ${\cal L}_D$, 
a part that involves four-fermion interactions, ${\cal L}_{\ell q}$, and a gluonic part, ${\cal L}_{G}$, 
\begin{equation}\label{Leff}
{\cal L}_{\rm eff}= {\cal L}_D + {\cal L}_{\ell q} + {\cal L}_{G} + ... .
\end{equation}
Here the ellipses denote effective operators that are not relevant for the following discussion. 
The dipole part in Eq.~(\ref{Leff}) is usually written as \cite{Hazard:2016fnc,Celis:2014asa}
\begin{eqnarray}\label{LD}
{\cal L}_{D} = -\frac{m_2}{\Lambda^2} \left[
\left( 
C_{DR}^{\ell_1\ell_2} \ \overline \ell_1 \sigma^{\mu\nu} P_L \ell_2 + 
C_{DL}^{\ell_1\ell_2} \ \overline \ell_1 \sigma^{\mu\nu} P_R \ell_2 
\right) F_{\mu\nu} + h.c. \right],
\end{eqnarray}
where $P_{\rm R,L}=(1\pm \gamma_5)/2$ is the right (left) chiral projection operator. The Wilson 
coefficients would, in general, be different for different leptons $\ell_i$. 

The four-fermion (dimension-six) lepton-quark Lagrangian is given by
\begin{eqnarray}\label{Llq}
{\cal L}_{\ell q} = -\frac{1}{\Lambda^2} \sum_q \Big[
\left( C_{VR}^{q\ell_1\ell_2} \ \overline\ell_1 \gamma^\mu P_R \ell_2 + 
C_{VL}^{q\ell_1\ell_2} \ \overline\ell_1 \gamma^\mu P_L \ell_2 \right) \ \overline q \gamma_\mu q &&
\nonumber \\
+ \
\left( C_{AR}^{q\ell_1\ell_2} \ \overline\ell_1 \gamma^\mu P_R \ell_2 + 
C_{AL}^{q\ell_1\ell_2} \ \overline\ell_1 \gamma^\mu P_L \ell_2 \right) \ \overline q \gamma_\mu \gamma_5 q &&
\nonumber \\
+ \
m_2 m_q G_F \left( C_{SR}^{q\ell_1\ell_2} \ \overline\ell_1 P_L \ell_2 + 
C_{SL}^{q\ell_1\ell_2} \ \overline\ell_1 P_R \ell_2 \right) \ \overline q q &&
\\
+ \
m_2 m_q G_F \left( C_{PR}^{q\ell_1\ell_2} \ \overline\ell_1 P_L \ell_2 + 
C_{PL}^{q\ell_1\ell_2} \ \overline\ell_1 P_R \ell_2 \right) \ \overline q \gamma_5 q 
\nonumber \\
+ \
m_2 m_q G_F \left( C_{TR}^{q\ell_1\ell_2} \ \overline\ell_1 \sigma^{\mu\nu} P_L \ell_2 + 
C_{TL}^{q\ell_1\ell_2} \ \overline\ell_1 \sigma^{\mu\nu} P_R \ell_2 \right) \ \overline q \sigma_{\mu\nu} q 
 &+& h.c. ~ \Big] .
\nonumber
\end{eqnarray}

The dimension seven gluonic operators can be either generated by some high scale physics or 
by integrating out heavy quark degrees of freedom \cite{Celis:2014asa,Petrov:2013vka},
\begin{eqnarray}\label{LG}
{\cal L}_{G} = -\frac{m_2 G_F}{\Lambda^2} \frac{\beta_L}{4\alpha_s} \Big[
\Big( C_{GR}^{\ell_1\ell_2} \ \overline\ell_1 P_L \ell_2 + 
C_{GL}^{\ell_1\ell_2} \ \overline\ell_1 P_R \ell_2 \Big)  G_{\mu\nu}^a G^{a \mu\nu} &&
\nonumber \\
+ ~ \Big( C_{\bar G R}^{\ell_1\ell_2} \ \overline\ell_1 P_L \ell_2 + 
C_{\bar G L}^{\ell_1\ell_2} \ \overline\ell_1 P_R \ell_2 \Big)  G_{\mu\nu}^a \widetilde G^{a \mu\nu}
 &+& h.c. \Big].
\end{eqnarray}
Here $\beta_L=-9 \alpha_s^2/(2\pi)$ is defined for the number of light active flavors, $L$, relevant to the scale 
of the process. All Wilson coefficients should also be calculated at the same scale.  
$\widetilde G^{a \mu\nu} = (1/2) \epsilon^{\mu\nu\alpha\beta} G^a_{\alpha\beta}$ is a dual to the
gluon field strength tensor.

The most general expression for the $V \to \ell_1 \overline \ell_2$ decay amplitude 
can be written as
\begin{eqnarray}\label{Spin1Amp}
{\cal A}(V\to \ell_1 \overline \ell_2) = \overline{u}(p_1, s_1) \left[
A_V^{\ell_1\ell_2} \gamma_\mu + B_V^{\ell_1\ell_2} \gamma_\mu \gamma_5 
+ \frac{C_V^{\ell_1\ell_2}}{m_{V}} (p_2-p_1)_\mu 
\right. ~~~~~~~~~~~~~~
\nonumber \\
\qquad + \left.
\frac{iD_V^{\ell_1\ell_2}}{m_{V} }(p_2-p_1)_\mu \gamma_5 \
\right] v(p_2,s_2) \ \epsilon^\mu(p).
\end{eqnarray}
where $A_V^{\ell_1\ell_2}$, $B_V^{\ell_1\ell_2}$, $C_V^{\ell_1\ell_2}$, and $D_V^{\ell_1\ell_2}$ are 
constants which depend on Wilson coefficients of the effective Lagrangian of Eq.~(\ref{Leff}) as well as on 
hadronic effects associated with meson-to-vacuum matrix elements (decay constants). The exact form of 
these form-factors is rather cumbersome and can be found in \cite{Hazard:2016fnc}. The amplitude of 
Eq.~(\ref{Spin1Amp}) leads to the branching fraction, which is convenient to represent as 
\begin{eqnarray}\label{BRSpin1}
\frac{{\cal B}(V \to \ell_1 \overline \ell_2)}{{\cal B}(V \to e^+e^-)} &=& 
\left(\frac{m_V \left(1-y^2\right)}{4\pi\alpha  f_V Q_q}\right)^2 \Big[ \left(\left|A_V^{\ell_1\ell_2}\right|^2 +  
\left|B_V^{\ell_1\ell_2}\right|^2\right)
+ \frac{1}{2} \left(1-2y^2\right)  \left(\left|C_V^{\ell_1\ell_2}\right|^2 +  \left|D_V^{\ell_1\ell_2}\right|^2\right) 
\nonumber \\
&+& y \ \mbox{Re}\left(A_V^{\ell_1\ell_2} C_V^{\ell_1\ell_2 *}+i B_V^{\ell_1\ell_2} D_V^{\ell_1\ell_2 *}\right) 
\Bigr] .
\end{eqnarray}
Here $y=m_{\ell_2}/m_V$. Comparing Eq.~(\ref{BRSpin1}) to experimental data one can constrain the Wilson coefficients of 
Eq.~(\ref{Llq}) that correspond to vector and tensor operators. They can be found in 
Table~\ref{tab1} \cite{Hazard:2016fnc,Abada:2015zea}.
\begin{table}
\begin{tabular}{ccccccc}
  & Leptons &\multicolumn{5}{c}{Initial state (quark)}\\
 \hline
Wilson coef, GeV$^{-2}$ & $\ell_1 \ell_2$ & $\Upsilon(1S) \ (b)$ & $\Upsilon(2S) \ (b)$ & $\Upsilon(3S) \ (b)$ 
 & $J/\psi \ (c)$ & $\phi \ (s)$  \\ \hline
$~$ & $\mu \tau$ & $ 5.6 \times 10^{-6}$ & $4.1 \times 10^{-6}$ & $3.5 \times 10^{-6}$ 
 & $5.5 \times 10^{-5}$ & n/a \\
$\left| C_{VL}^{q\ell_1\ell_2}/\Lambda^2 \right|$ & $e \tau$ & $-$ & $4.1 \times 10^{-6}$ & $4.1 \times 10^{-6}$ 
 & $1.1 \times 10^{-4}$ & n/a \\ 
$~$ & $e \mu$ & $-$ & $-$ & $-$  
& $1.0 \times 10^{-5}$ & $2 \times 10^{-3}$  \\
\hline
$~$ & $\mu \tau$  & $ 5.6 \times 10^{-6}$ & $4.1 \times 10^{-6}$ & $3.5 \times 10^{-6}$ 
 & $5.5 \times 10^{-5}$ & n/a \\
$\left| C_{VR}^{q\ell_1\ell_2}/{\Lambda^2} \right|$ & $e \tau$ & $-$ & $4.1 \times 10^{-6}$ & $4.1 \times 10^{-6}$ 
 & $1.1 \times 10^{-4}$ & n/a \\
$~$ & $e \mu$ & $-$ & $-$ & $-$ 
 & $1.0 \times 10^{-5}$ & $2 \times 10^{-3}$ \\
\hline
$~$  & $\mu \tau$  & $ 4.4 \times 10^{-2}$ & $3.2 \times 10^{-2}$ & $2.8 \times 10^{-2}$ 
 & $1.2$ & n/a \\
$\left| {C_{TL}^{q\ell_1\ell_2}}/{\Lambda^2} \right|$ & $e \tau$ & $-$ & $3.3 \times 10^{-2}$ & $3.2 \times 10^{-2}$ 
 & $2.4$ & n/a \\
$~$ & $e \mu$ & $-$ & $-$ & $-$ 
 & $4.8$ & $1 \times 10^{4}$ \\
\hline
$~$  & $\mu \tau$  & $ 4.4 \times 10^{-2}$ & $3.2 \times 10^{-2}$ & $2.8 \times 10^{-2}$ 
 & $1.2$ & n/a \\
$\left| {C_{TR}^{q\ell_1\ell_2}}/{\Lambda^2} \right|$ & $e \tau$ & $-$ & $3.3 \times 10^{-2}$ & $3.2 \times 10^{-2}$ 
 & $2.4$ & n/a \\
$~$ & $e \mu$ & $-$ & $-$ & $-$ 
 & $4.8$ & $1 \times 10^{4}$ \\
\hline
\end{tabular}
\caption{Constraints on the Wilson coefficients of four-fermion operators from $1^{--}$ quarkonium decays. Dashes signify the 
absence of experimental data; ``n/a" means that the transition is forbidden by phase space.}
\label{tab1}
\end{table}

Similar analysis can be performed for LFV decays of scalar ($S$) and pseudoscalar ($P$) quarkonium states. 
The most general expression for the $S/P \to \ell_1 \overline \ell_2$ decay amplitude is
\begin{eqnarray}\label{Spin0Amp}
{\cal A}(S/P\to \ell_1 \overline \ell_2) = \overline{u}(p_1, s_1) \left[
E_{S/P}^{\ell_1\ell_2}  + i F_{S/P}^{\ell_1\ell_2} \gamma_5 
\right] v(p_2,s_2) \,
\end{eqnarray}
where, as in the case of the vector quarkonium decays $E_{S/P}^{\ell_1\ell_2} $ and $F_{S/P}^{\ell_1\ell_2}$  are 
dimensionless constants which depend on the underlying Wilson coefficients of the effective 
Lagrangian of Eq.~(\ref{Leff}) and on decay constants.

The amplitude of Eq.~(\ref{Spin0Amp}) leads to the branching ratio for flavor off-diagonal leptonic 
decays of pseudoscalar mesons:
\begin{eqnarray}\label{BRSpin0}
{\cal B}(S/P \to \ell_1 \overline \ell_2) = \frac{m_{S/P}}{8\pi \Gamma_{S/P}} \left(1-y^2\right)^2
\left[\left|E_{S/P}^{\ell_1\ell_2}\right|^2 + \left|F_{S/P}^{\ell_1\ell_2}\right|^2\right].
\end{eqnarray}
Here $\Gamma_{S/P}$ is the total width of the scalar or pseudoscalar state and $y=m_{\ell_2}/m_{S/P}$. 
It is interesting to note that scalar quarkonium decays are mostly sensitive to the scalar operators in 
Eq.~(\ref{Llq}), eliminating the need an assumption of single operator dominance \cite{Hazard:2016fnc}.

The decays of the scalar $S=\chi_{q0}$ or pseudoscalar $P=\eta_q$ states are difficult to study at colliders. 
However, the following trick could be employed. Since vector states are abundantly produced both
at $e^+e^-$ and hadronic machines, a resonant two-body radiative transitions of vector states 
\begin{equation}
{\cal B}(V \to \gamma  \ell_1 \overline \ell_2) = 
{\cal B}(V \to \gamma \ S/P) {\cal B}(S/P \to \ell_1 \overline \ell_2),
\end{equation} 
could be used to produce scalar $S$ and/or pseudoscalar $P$ states. The branching ratios 
for the radiative transitions $V \to \gamma \ S/P$ are rather large, e.g. 
\begin{eqnarray}\label{BranchRad}
&&  {\cal B}(\psi(2S) \to \gamma \chi_{c0} (1P)) = 9.99 \pm 0.27\% \ ,
\nonumber \\
&&  {\cal B}(\Upsilon(3S) \to \gamma \chi_{b0} (2P)) = 5.9 \pm 0.6\% \ .
\nonumber \\
&& {\cal B}(J/\psi \to \gamma \eta_c) = 1.7 \pm 0.4\% \ ,
\end{eqnarray}
An estimate \cite{Godfrey:2015vda} shows that with the integrated luminosity of 
${\cal L}=250$ fb$^{-1}$ the number of produced $\chi_b$ states could reach tens of millions,
making such studies very feasible.

%%%%%%%%%%%%%%%%%%%%%
\section{Probing rare leptonic charm transitions in production experiments}\label{Section4}

As was mentioned in Sect. \ref{Section2}, the rate of the simplest FCNC decay, $D^0 \to \ell^+ \ell^-$,
is suppressed by the helicity, i.e. by the mass of the final state leptons. This could complicate studies of lepton 
flavor universality as the branching ratio of $D^0 \to e^+ e^-$ is tiny. Experimental studies of a similar transition, 
$D^* \to e^+e^-$, that is not helicity suppressed, are not feasible as $D^*$, contrary to the $D^0$, also 
decays strongly or electromagnetically with much larger rates.  

%%%%%%%%%%%%%%%%%%%%%%%%%%%%%%%%%%%
\begin{figure}[t]
\begin{center}
\includegraphics[width=6cm]{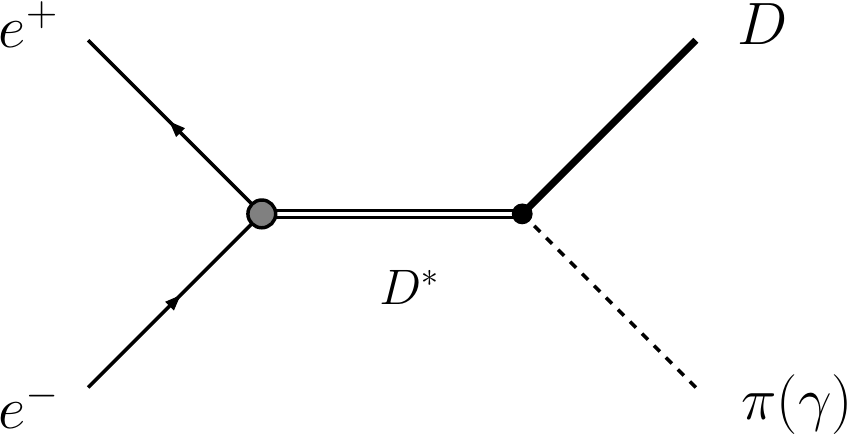}
\end{center}
\caption{\label{figure} Probing the $ c\bar{u}\to e^+ e^-$ vertex
with the $D^*(2007)^0$ resonance production in $e^+e^-$ collisions.}
\end{figure}
%%%%%%%%%%%%%%%%%%%%%%%%%%%%%%%%%%%

An interesting alternative to studies of $D^*$ decays is to measure the corresponding {\it production} process 
$e^+e^- \to D^*$, as shown in Fig.~\ref{figure} \cite{Khodjamirian:2015dda}. This is possible at an $e^+e^-$ collider, 
such as BEPCII or VEPP-2000, tuned to run at the center-of-mass energy corresponding to the mass of the $D^*$ meson, 
$\sqrt{s} \approx 2007$ MeV. The produced $D^{*0}$ resonance, tagged by a single charmed particle in the final state,
will decay strongly ($D^{*0}\to D^0\pi^0$) or electromagnetically ($D^{*0}\to D^0\gamma$) with branching fractions of 
$(61.9\pm 2.9)\%$ and $(38.1\pm 2.9)\%$ respectively. This process, albeit very rare, has clear advantages 
for NP studies compared to the $D^0 \to e^+e^-$ decay: the helicity suppression is absent, and a richer set of effective 
operators can be probed. It is also interesting to note that contrary to other rare decays of charmed mesons, long-distance 
SM contributions are under theoretical control and contribute at the same order of magnitude as the short-distance ones. 
Similar opportunities exist for $B$-decays as well \cite{Khodjamirian:2015dda,Grinstein:2015aua}.

%%%%%%%%%%%%%%%%%%%%%
\section{Conclusions}\label{Section5}

The apparent absence of any hints of new particles from current direct searches at Large Hadron Collider (LHC) experiments 
makes careful studies of their possible quantum effects an important tool in our arsenal of methods for probing physics
beyond the SM. Abundance of charm data in the current and future low energy flavor experiments makes 
it possible to study NP in QFV and LFV rare decays of charmed mesons with ever increased precision. The obtained constraints 
from a variety of methods described in this talk are competitive with the bounds obtained from the continuing direct searches 
for NP particles at the LHC.

%%%%%%%%%%%%%%%%%%%%%

\end{document}